\documentclass[prl,twocolumn]{revtex4}

\usepackage{enumerate}
\usepackage{amsfonts,amssymb,amsmath}
\usepackage[]{graphics,graphicx,epsfig}
\usepackage{amsthm}
\usepackage{float}

\usepackage{graphicx}
\usepackage{xcolor}
\usepackage{qcircuit}

\usepackage{tikz}
\usetikzlibrary{arrows}

\usepackage{pdfpages}
\usepackage{epstopdf}
\DeclareGraphicsExtensions{.png,.pdf}

\usepackage{graphicx}
\usepackage{dcolumn}
\usepackage{natbib}
\usepackage{color}
\usepackage{multirow}
\usepackage{ulem}
\newcommand{\ket}[1]{|{#1}\rangle}
\newcommand{\bra}[1]{\langle{#1}|}
\usepackage{array}
\newcolumntype{P}[1]{>{\centering\arraybackslash}p{#1}}

\begin{document}


\title{Quantum Resetting Protocols Based on Synchronizing Words }


\author{J\c{e}drzej Stempin}
\affiliation{Institute of Spintronics and Quantum Information, Faculty of Physics, Adam Mickiewicz University, 61-614 Pozna\'n, Poland}

\author{Jan W{\'o}jcik}
\affiliation{Institute of Spintronics and Quantum Information, Faculty of Physics, Adam Mickiewicz University, 61-614 Pozna\'n, Poland}

\author{Andrzej Grudka}
\affiliation{Institute of Spintronics and Quantum Information, Faculty of Physics, Adam Mickiewicz University, 61-614 Pozna\'n, Poland} 

\author{Marcin Karczewski}
\affiliation{Institute of Spintronics and Quantum Information, Faculty of Physics, Adam Mickiewicz University, 61-614 Pozna\'n, Poland}

\author{Pawe{\l} Kurzy{\'n}ski}
\email{pawel.kurzynski@amu.edu.pl}
\affiliation{Institute of Spintronics and Quantum Information, Faculty of Physics, Adam Mickiewicz University, 61-614 Pozna\'n, Poland}

\author{Antoni W{\'o}jcik}
\affiliation{Institute of Spintronics and Quantum Information, Faculty of Physics, Adam Mickiewicz University, 61-614 Pozna\'n, Poland}

\date{\today}


\begin{abstract}
Resetting a system's state plays a fundamental role in physics, engineering, computer science, and many other fields. Here we focus on a method originally proposed in automata theory. The state of an automaton evolves according to a set of rules. These rules form an alphabet, and one can apply these rules one after another. Such a sequence of rules is known as a word. Some particular words, known as synchronizing words, enable a system to evolve into a predetermined state regardless of its initial configuration. This process, which is inherently irreversible, appears at first glance to be incompatible with the unitarity of quantum mechanics. Unitary evolution preserves information, hence forbids such classical resetting to a predetermined state. In this work, we introduce a novel resetting protocol based on quantum-synchronizing words by incorporating auxiliary qubits whose states encode rules of the automaton's alphabet. We further propose a quantum circuit that realizes this protocol and can be implemented on a quantum computer. Additionally, we establish a connection between this approach and Kraus channels, showing that quantum synchronizing words can be achieved without explicit reference to the states of ancillary qubits. Our results bridge classical and quantum notions of synchronizing words, shedding light on the interplay between quantum information processing and non-unitary dynamics.
\end{abstract}

\maketitle

\section{Introduction}
Synchronization is a broad concept that appears in various fields of science. In physics it is mostly considered to be a phenomenon in which two, or more, oscillators evolve towards the same frequency and phase. However, in computer science, particularly in automata theory, synchronization is a process that takes an automaton to a predetermined state, regardless of its initial state \cite{vcerny1964poznamka,jurgensen2008synchronization}. Despite apparent differences, both notions of synchronization have one important common feature: it is a process that contracts the allowable state space of the system.

To illustrate automata synchronization, consider a highly secure bank safe with a unique locking mechanism. The safe unlocks only when the correct numerical code, ranging from $0$ to $9$, is entered. However, entering an incorrect number results in irreversible destruction of its contents. The safe is operated using three buttons: $a$, $b$, and "Enter". Internally, the system starts from an unknown initial state, some predetermined number, and pressing $a$ or $b$ applies deterministic transformations to this state. While the functions of $a$ and $b$ are publicly known, the lack of knowledge about the initial state makes it extremely difficult to determine the precise sequence of operations required to reach the correct code. As a result, even though both the transformations and the final password are public, opening the safe without triggering destruction remains a formidable challenge.

This problem underscores the importance of a synchronization protocol—an algorithmic sequence of $a$ and $b$ that ensures convergence to a unique final state, irrespective of the initial condition. Such sequences, known as \textit{synchronizing words} \cite{ryzhikov2020synchronizing,van2022synchronizing,vcerny1964poznamka,eppstein1990reset,jurgensen2008synchronization}, are central to automata theory and have broad applications in control theory, coding, and symbolic dynamics. A well-designed synchronization protocol would allow the safe to be opened reliably, bypassing the uncertainties of the initial state.

In this paper, we extend the concept of  synchronizing words to quantum systems. A natural first approach is to translate classical synchronization schemes into the language of unitary quantum dynamics. However, a fundamental challenge arises: for synchronization to occur, there must exist operations that map multiple initial states to the same final state. This requirement conflicts with the reversibility of unitary evolution, making direct translation impossible.

To overcome this limitation, we propose a method that achieves reset of quantum automata by introducing auxiliary subsystems. We demonstrate how this approach can be implemented on a quantum computer for an arbitrary number of states. Furthermore, we explore an alternative framework based on noisy Kraus channels, which enables reset without explicit reference to the states of additional subsystems.  This approach naturally extends our previously proposed concept \cite{grudka2025quantum} and establishes a robust foundation for the development of effective and universal quantum computation tools.

\section{Synchronizing words in classical Automaton Theory}


A Deterministic Finite Automaton (DFA) is formally defined as a 5-tuple $(Q, \Sigma, \delta, F, q_0)$, where $Q$ represents a non-empty set of possible states, $\Sigma$ denotes the input alphabet, $\delta: Q \times \Sigma \rightarrow Q $
is the transition function governing system's dynamics, $F \subseteq Q$ is the set of accepting states, and $q_0 \in Q$ is the initial state \cite{NAEEMIRFAN201389}. The whole language over the alphabet $\Sigma$ is denoted as $\Sigma^{*}$. A DFA can be represented as a collection of directed graphs, where each graph corresponds to the action of function $\delta_a: Q \rightarrow Q$ defined as $\delta_a(q) = \delta(q, a)$ for a specific symbol $a \in \Sigma$ and any $q\in Q$. Given a transition function $\delta$ and an input word (a sequence of symbols from $\Sigma$), one can determine the resulting state of the DFA. A word $\bar{w} \in \Sigma^{*}$ that drives the automaton to the same final state regardless of the initial state is termed a synchronizing word. Several algorithms have been developed for identifying synchronizing words for a given DFA \cite{ryzhikov2020synchronizing,van2022synchronizing,vcerny1964poznamka,eppstein1990reset,jurgensen2008synchronization}.

For DFAs consisting solely of cycles, synchronizing words do not exist, as synchronization/reset requires the presence of contractions, namely nodes where multiple edges converge. This feature is crucial for irreversibility of the dynamics. Consequently, we focus our analysis on modified cycles featuring an additional node connected to a main cycle. A fundamental case involves a DFA with two states and two input symbols, as illustrated in Figure 1.

\begin{figure}[h!]
  
     \includegraphics[width=0.6\linewidth]{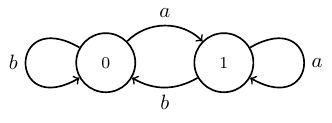}
        \caption{Trivial example of DFA composed of $Q = \{0,1\}$ and $\Sigma = \{a,b\}$. For each case of initial position, word composed of a single letter is a synchronizing word}
    \label{fig:main}
\end{figure}

This basic model can be generalized by incorporating additional nodes into the external loop. We establish the graph corresponding to symbol $a$ as our reference, with nodes labeled in ascending order. The graph for symbol $b$ can be derived by applying a permutation $\pi$ to the reference graph. Both graphs are depicted in Figure \ref{cycle}. For analytical simplicity, we assume $\pi_0 = 1$ and $\pi_1 = 0$. For the sake of simplicity, we denote permutation for which $\pi_{0} = 1$, $\pi_{1} = 0$ and $\pi_{k}=k$ for $k \geq 2$ as a transposition $\pi = (1,0)$. In this elementary case where $\pi = (1,0)$, the shortest synchronizing word $\bar{w}$ is expressed by:
\begin{equation}
\bar{w} \equiv a^{(n-1)\text{mod} \hspace{0.1cm}2}(ab)^{\lfloor(n-1)/2\rfloor}
\end{equation}
where $n$ is the number of nodes. 
We adopt the convention of reading words from right to left, aligning with the standard order of quantum operations. This approach ensures consistency with the mathematical formalism governing quantum transformations. This synchronizing word invariably directs the DFA to state $1$. Conversely, interchanging $a$ and $b$ yields a synchronizing word that terminates in state $0$. Notably, in this example, the length of the synchronizing word is precisely $n-1$.

\begin{figure}[h!]
         \includegraphics[width=0.65\linewidth]{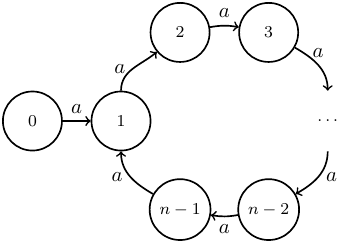}
         \includegraphics[width=0.65\linewidth]{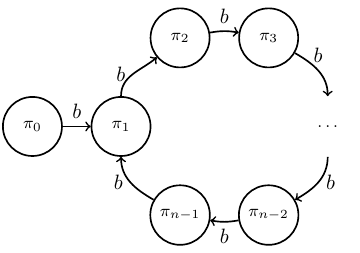}
        \caption{Graph corresponding to the letter $a$ (upper one), and corresponding to the letter $b$ (lower one), for the general case of considered systems.}
    \label{n_cycles}
\end{figure}

\section{Unitary protocol for synchronizing words}
Numerous techniques have been introduced for the implementation of quantum cellular automata \cite{schumacher2004reversible,perez2007local,wiesner2008quantum,grossing1988quantum,farrelly2020review,moore2000quantum}. In this paper, we propose a novel quantization approach for the DFA that maintains maximal compatibility with the concept of synchronizing words. 

Evolution on directed graphs can be difficult to quantize due to the presence of contractions. However, this limitation can be overcome through the addition of an auxiliary control system. We propose a formalism where states of the DFA are described by vectors in a n-dimnesional Hilbert space $\mathcal{H}_{s}$. Individual letters from the alphabet are encoded by vectors from qubit space. Consequently, the total state of a k-letter word and corresponding nodes is encoded by a vector belonging to the tensor product space $\mathcal{H}_{2}^{\otimes k}\otimes \mathcal{H}_{s}$

In our model, at any given timestep, the system interacts only with a particular qubit $j$. This interaction is described by a unitary operator $U_{j}$ which is defined according to the permutation rule written in eq. \ref{cycle}.

\begin{equation}
    \begin{array}{ccc}
      \ket{a}_{j}\ket{0} & \leftarrow  & \ket{b}_{j}\ket{\pi_{n-1}}\\
      \downarrow &  & \uparrow \\
       \ket{a}_{j}\ket{1} &  & \ldots\\
      \downarrow & & \uparrow\\
       \ket{a}_{j}\ket{2} & & \ket{b}_{j}\ket{\pi_{2}} \\
      \downarrow & & \uparrow\\
      \ldots & & \ket{b}_{j}\ket{0} \\
      \downarrow & & \uparrow\\
      \ket{a}_{j}\ket{n-1} & \rightarrow & \ket{b}_{j}\ket{1} \\
    \end{array}
    \label{cycle}
\end{equation}
For the specific case, discussed in previous section, where $\pi = (0,1)$, starting from an arbitrary superposition of graph states, one can determine the final state of the system by applying the sequence of unitary operations as defined in eq. \ref{cycle}. The transformation for $n=4$ nodes proceeds as follows:
\begin{equation}
\begin{array}{c}
   U_{3}U_{2}U_{1} \ket{a}_{3}\ket{b}_{2}\ket{a}_{1} \otimes (\alpha \ket{0} + \beta \ket{1} + \gamma \ket{2} + \delta \ket{3})=      \\
   \\
      =(\alpha \ket{aba} + \beta\ket{bba} + \gamma\ket{aaa} + \delta\ket{abb})\otimes \ket{1}
\end{array}
\end{equation}

\section{Sychronizing quantum circuit}

Let's consider graphs with $n$ nodes, where $n$ is a power of $2$. For practical quantum implementation, we encode the nodes in states of multiple qubits $\{g_{1},\ldots,g_{m}\}$, where $m = \log_{2}{n}$ qubits are required to represent $n$ nodes. Additionally, we introduce controlling qubits $\{q_{j}\}_{j=0}^{n-1}$, with each step $j$ involving interaction between the graph qubits and only one controlling qubit $q_{j}$. The total gate implements the operator $U_{j}$ defined in eq. \ref{n_cycles}. For notational convenience, we identify $\ket{a}$ with qubit's state $\ket{0}$ and $\ket{b}$ with $\ket{1}$.

\begin{figure}[!h]
    \includegraphics[width=0.85\linewidth]{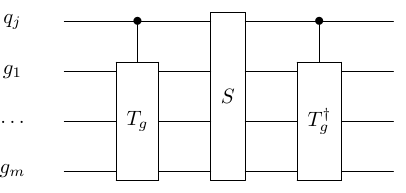}
    \caption{ Circuit corresponding to one step of the dynamics involving interaction with only one auxiliary qubit.}
    \label{full_circ}
\end{figure}

\begin{figure}[!h]
    \centering
    \includegraphics[width=0.5\linewidth]{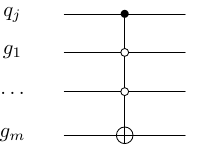}
    \caption{Total $T$ gate for the case of $\pi = (1,0)$.}
    \label{Tgate}
\end{figure}

The proposed quantum circuit, depicted in Figure \ref{full_circ}, consists of three essential components listed below.
\begin{enumerate}
    \item \textbf{Basis transformation Operator} $T$. This operator transforms input vectors to a basis in which evolution manifests itself as a simple shift on a $2m$-dimensional space composed of position space and an auxiliary qubit's space. It requires conditional operation that maps nodes indexed in sequential numerical order to nodes reindexed according to the permutation mapping $\pi$ when the control qubit is in the state $\ket{1}$ of letter $b$. This implementation can be realized through application of:
    \begin{equation}
        T = \ket{0}\bra{0}\otimes I + \ket{1}\bra{1} \otimes T_{g}
    \end{equation}
    where:
    \begin{equation}
        T_{g} = \sum_{i = 1}^{m} \ket{\pi_{i}}\bra{i}
    \end{equation}
    In Figure \ref{Tgate}, we present the $T$ gate for the simplest case of $\pi = (1,0)$.
    
    \item \textbf{Shift operator} $S$. This operator implements cyclic index incrementation, which is essential for maintaining the proper evolution dynamics according to rule from eq. \ref{cycle}.
    \begin{equation}
        S = \sum_{i = 1}^{2m} \ket{(i+1)\hspace{0.1cm} \text{mod} \hspace{0.1cm} 2m}\bra{i}
    \end{equation}
    \item \textbf{Inverse transformation} $T^{\dagger}$. This operator rotates back to the computational basis in which the binary representation of a certain node is in accordance with the primal setup.
\end{enumerate}

Each of the gates described above can be constructed through a series of swap operations between selected states. This approach is valid because any permutation can be decomposed into a product of $2$-cycles (transpositions). For example, the operator $T$ in the case where $\pi = (1,0)$ takes the form of a multi-controlled $CNOT$ gate depicted in Figure \ref{Tgate}.

The same methodology can be applied to derive a decomposition of the shift operator. It's easy to show that the shift permutation can be expressed as a product of $2$-cycles in the following form:
\begin{equation}
\begin{array}{cc}
    (2m-1,0,\ldots,2m-2) =     \\
   =  (2m-1,0)(2m-1,1)\ldots(2m-1,2m-2) 
\end{array}
\end{equation}

\begin{figure}[!h]
    \centering
    \includegraphics[width=0.7\linewidth]{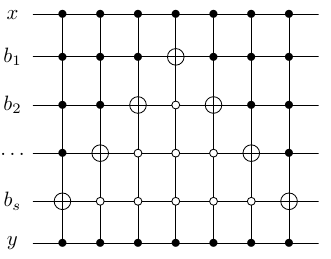}
    \caption{General swap between states $k$ and $2m$ where $b_{1},\ldots,b_{s}$ represent bits valued as $0$ in binary representation of $k$ and $x,y$ represent all bits valued as $1$. The order of qubits in this circuit has been changed in order to resemble the underlying structure in more concise way.}
    \label{general_swap}
\end{figure}

Therefore, one should apply the sequence of transpositions between the the highest valued state $2m-1$ and certain number $k$, chosen in a particular step. Next, one can use the fact that the binary representation of $2m-1$ is a simple sequence of ones of length $\log_{2}(2m)$.  Using this fact, each transposition should be constructed according to the following recipe: if $b_{1},\ldots,b_{s}$ denote bits on which the binary representation of state $k$ has $0$s (differs from the binary representation of $2m-1$) and $x,y$ denote bits that take the same value for both $k$ and $2m-1$ one should apply the general swap gate constructed from a sequence of swaps followed by the reversed sequence of operations. As a result, only the states $\ket{k}$ and $\ket{2m-1}$ are affected by this operation with effect that $\ket{k}$ is assigned to $\ket{2m-1}$ and {\it{vice versa}}.

The whole circuit should be applied separately for each controlling qubit. Assuming that qubits $\{q_{j}\}$ were initialized according to the letters of the synchronizing word, one will obtain a synchronized state on the graph qubits $\{g_{1},\ldots,g_{m}\}$ regardless of their initial state configuration. Detailed information about possible implementation on quantum computer can be found in Appendix A.

\section{Quantum walk interpretation}
For the special case of a totally reversed permutation on nodes from $2$ to $n-1$ which is $\pi = (1,0)(n-1,\ldots,2)$, we can interpret the resetting circuit as a Quantum Random Walk with a modified step operator. This perspective provides valuable insights into the quantum dynamics underlying the resetting process.

The step operator $S$ governing the quantum walk is defined as:
\begin{equation}
\left\{
\begin{array}{ll}
    S\ket{a}\ket{x} = \ket{a}\ket{x + 1}, & \text{for } x \neq n-1, \\
    S\ket{a}\ket{n-1} = \ket{b}\ket{1}, & \\
    S\ket{b}\ket{x} = \ket{b}\ket{x - 1}, & \text{for } x \neq 0,x\neq 2 \\
    S\ket{b}\ket{0} = \ket{b}\ket{n-1} & \\
    S\ket{b}\ket{2} = \ket{a}\ket{0} & \\
\end{array}
\right.
\label{QwalkS}
\end{equation}

Under these transition rules, every initial state $\ket{\psi_{0}}$ of the system, prepared with ancillary coin states:
\begin{equation}
   \ket{\Psi_{0}} = \ket{a}\ket{ab}^{\otimes (n-3)} \otimes \ket{\psi_{0}}
\end{equation}
will be driven to the state $\ket{1}$ after sequence of steps applied to each ancillary qubit. Furthermore, this representation establishes a direct connection between quantum automata theory and the well-studied field of quantum walks, potentially allowing techniques from one domain to be applied to problems in the other \cite{lovett2010universal, costa2018quantum,huerta2020quantum,kempe2009quantum}. The effectiveness of this resetting protocol scales with the dimension of the state space, requiring $\mathcal{O}(n)$ operations to achieve complete resetting.

\begin{figure}[!h]
    \includegraphics[width=1\linewidth]{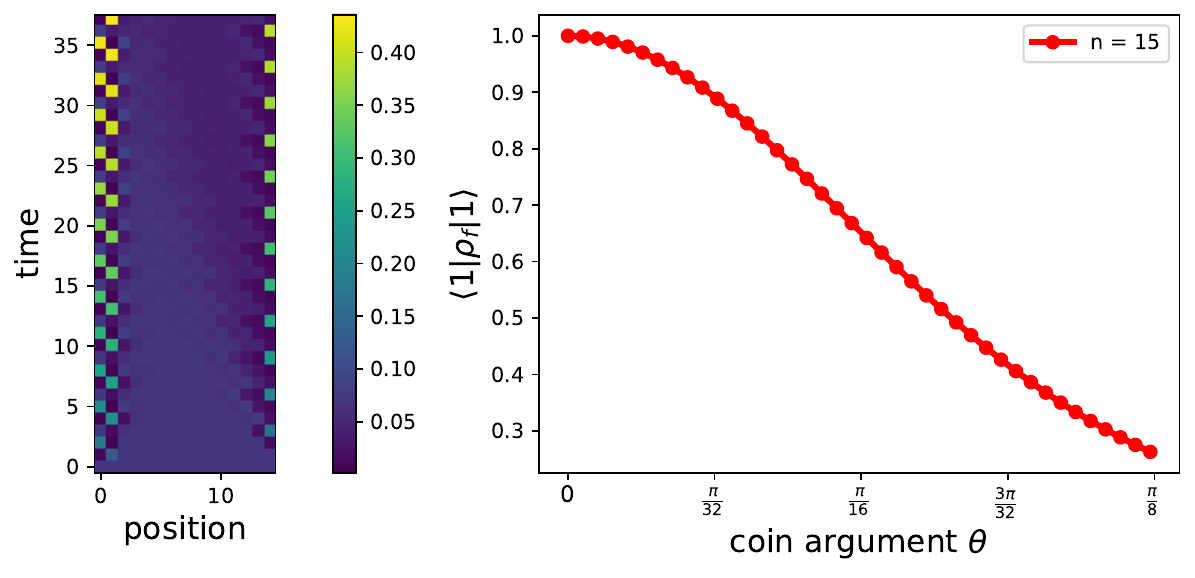}
    \caption{On the left: probability distribution over time in position domain of Quantum Walk with resetting step operator defined in equation \ref{QwalkS} and value of $\theta = \frac{\pi}{11}$. On the right: fidelity of the final state $\rho_{f}$ after the evolution as a function of coin operator argument $\theta$. Initial state was chosen as equally weighted superposition of position states.}
    \label{QWalkIm}
\end{figure}

In Figure \ref{QWalkIm} we present the simulation of the Quantum Walk with resetting step operator. Total evolution of the system is given by the product:
\begin{equation}
    U = SC_{\theta}
\end{equation}
where $C_{\theta} = \exp{(-i\sigma_{y}\theta)}$ is the standard Quantum Walk coin operator.  The success of reset started from initially  delocalized state strongly depends on the value of $\theta$. Nevertheless, in the small range of parameter's value, between $0$ and $\pi/32$, the protocol is robust against perturbations and the final state is still a good approximation of desired state. The functional relationship of $\langle 1|\rho_{f}|1\rangle$ manifests invariance with respect to the dimensionality of the position space. This independence emerges from a fundamental insight: the quantum particle's dynamics experiences significant perturbation from coin tossing exclusively at the node where contraction materializes. Consequently, the global extent of the graph remains inconsequential to this particular quantum phenomenon.

\section{Resetting with noisy channels}

The reset protocol can be generalized beyond explicit reference to the controlling system's state through the formalism of noisy, quantum channels \cite{nielsen2010quantum}. We define a general rotation operation parametrized by the angle $\varphi$:
\begin{equation}
\begin{array}{c}
 R_{ij}^{[k]}(\varphi) =  \sum_{p\neq i,j,k} |p \rangle\langle p| + \cos{\varphi}|i\rangle \langle i |  + \\
 \\
 - \sin{\varphi} |i \rangle\langle j | + \sin{\varphi} |j \rangle \langle i | + \cos{\varphi} 
 |j\rangle\langle j| \\
\end{array}
\end{equation}
This operator is not a valid quantum operation because it doesn't include the state indexed by $k$, therefore it should be always completed with the second operation, which dictates what to do with state $\ket{k}$. In particular, when $\varphi = \pi / 2$, operator $R_{ij}^{[k]}(\varphi)$ implements a $NOT$ gate with an accompanying relative phase shift by $\pi$ between states $\ket{i}$ and $\ket{j}$. 
One more time, the decomposition of permutations into the product of transpositions will be used to obtain the complete Kraus channel. We define two operators:
\begin{equation}
    A_{1}(\varphi_{A}) = R_{12}^{[0]}(\varphi_{A})R_{23}^{[0]}(\varphi_{A}) \ldots R_{n-2,n-1}^{[0]}(\varphi_{A})
\end{equation}
\begin{equation}
    B_{1}(\varphi_{B}) = R_{02}^{[1]}(\varphi_{B})R_{23}^{[1]}(\varphi_{B}) \ldots R_{n-2,n-1}^{[1]}(\varphi_{B})
\end{equation}
Those two operations correspond to the great cycle of graphs from Figure \ref{cycle}. In order to resemble the structure of the noisy channels, one should also include two additional operations:
\begin{equation}
    A_{2} = |1\rangle\langle 0| \hspace{1cm} B_{2} = | 0 \rangle\langle 1| 
\end{equation}
Finally, the net effect of each Kraus channel on the density matrix is given by:
\begin{equation}
    A(\rho) = A_{1}\rho A_{1}^{\dag} + A_{2}\rho A_{2}^{\dag}
\end{equation}
\begin{equation}
    B(\rho) = B_{1}\rho B_{1}^{\dag} + B_{2}\rho B_{2}^{\dag}
\end{equation}

\begin{figure}[!h]
    \includegraphics[width=1\linewidth]{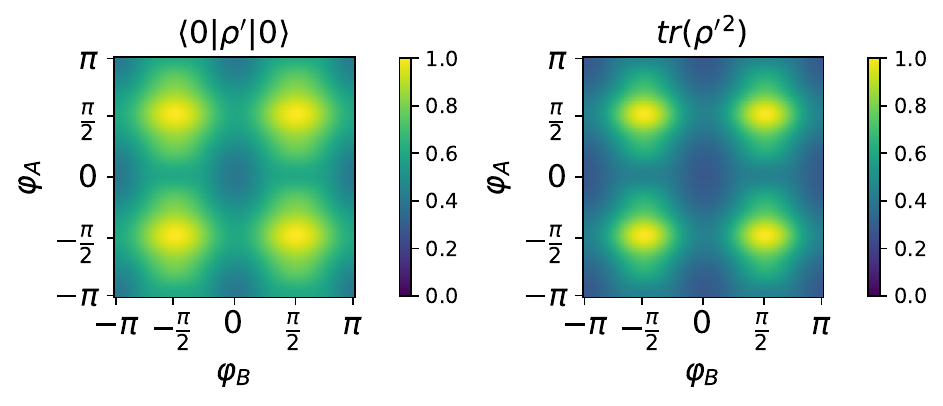}
    \includegraphics[width=1\linewidth]{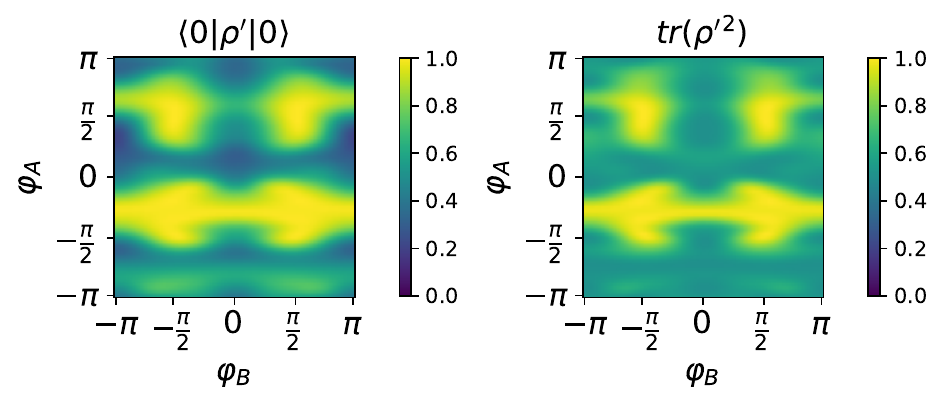}
    \caption{Fidelity for $\ket{0}$ state (on the left) and purity (on the right) for the graph with $n = 5$ nodes and the resulting state  after the protocol $\rho' = ABAB(\rho)$. The upper maps are calculated for the initially mixed state $\rho = \frac{1}{n}I$ and the lower maps are calculated for initial pure state constructed from equally weighted superposition.}
    \label{fidelity}
\end{figure}

Figure \ref{fidelity} presents the quantitative analysis of the system's response to the complete synchronizing word sequence. The figure displays both the fidelity (left panel) and purity (right panel) of the final quantum state as functions of the rotation angles. These results were systematically computed across a parametric space of varying $\varphi_A$ and $\varphi_B$ values. The resultant heat maps demonstrate that the reset protocol maintains significant robustness against minor perturbations that induce stochastic information degradation. This resilience suggests that the reset mechanism possesses inherent stability properties even in non-ideal quantum information processing environments where noise-induced decoherence effects are present.

\section{Conclusions}

In this paper, we have presented a novel approach to constructing a specified subclass of quantum cellular automata. We have demonstrated that this construction can be achieved through the exploitation of ancillary systems—specifically qubits—thereby establishing a meaningful connection between our framework and quantum information processing paradigms. The protocol we propose enables the deterministic reset of a quantum system to a specifically chosen state from an arbitrary initial superposition.

We have developed two methodological approaches: the first explicitly leverages the structure of the auxiliary environment space, for which we have designed a corresponding quantum circuit implementation; the second operates within the framework of noisy channels with explicit decoherence processes. This dual perspective facilitates the integration of our work with diverse theoretical models of generalized quantum dynamics. The Kraus channel formalism is particularly noteworthy, which provides an intriguing connection to our previous work \cite{grudka2025quantum}, in which we proposed reset protocols as a methodology to build a universal quantum computing framework for qutrit states. The arguments presented herein can be interpreted as a natural extension of these ideas, with significant potential for realizing a comprehensive set of gates capable of steering arbitrary $n$-dimensional quantum states to any target configuration.

Furthermore, we have demonstrated the implementation of our protocol within the well-established structure of Quantum Random Walks. In this context, we have shown that while the introduction of a coin operator perturbs the resetting effect, there exists a well-defined parameter regime within which these perturbations remain inconsequential to the desired outcome. In light of these findings, our protocol emerges as a versatile and adaptable tool applicable to a broad spectrum of quantum systems, with implications spanning fundamental quantum mechanics, quantum information science, and quantum computing architectures.

\vspace{1cm}
{\it Acknowledgements.} This research is supported by the Polish National Science Centre (NCN) under the Maestro Grant no. DEC-2019/34/A/ST2/00081.

\appendix
\section{Appendinx A: supplementary materials}
All relevant calculations and numerical results, including the implementation of quantum circuit described in this manuscript, have been made publicly available in our GitHub repository: \url{https://github.com/JedrekSt/Quantum_Synchronization_Protocol/tree/main}. The repository contains comprehensive documentation of the computational methods employed across multiple platforms, simulation data, and source code to facilitate reproducibility and further investigation of the resetting mechanisms presented herein.

\bibliographystyle{apsrev4-2}
\bibliography{main}

\begin{thebibliography}{18}%
\makeatletter
\providecommand \@ifxundefined [1]{%
 \@ifx{#1\undefined}
}%
\providecommand \@ifnum [1]{%
 \ifnum #1\expandafter \@firstoftwo
 \else \expandafter \@secondoftwo
 \fi
}%
\providecommand \@ifx [1]{%
 \ifx #1\expandafter \@firstoftwo
 \else \expandafter \@secondoftwo
 \fi
}%
\providecommand \natexlab [1]{#1}%
\providecommand \enquote  [1]{``#1''}%
\providecommand \bibnamefont  [1]{#1}%
\providecommand \bibfnamefont [1]{#1}%
\providecommand \citenamefont [1]{#1}%
\providecommand \href@noop [0]{\@secondoftwo}%
\providecommand \href [0]{\begingroup \@sanitize@url \@href}%
\providecommand \@href[1]{\@@startlink{#1}\@@href}%
\providecommand \@@href[1]{\endgroup#1\@@endlink}%
\providecommand \@sanitize@url [0]{\catcode `\\12\catcode `\$12\catcode `\&12\catcode `\#12\catcode `\^12\catcode `\_12\catcode `\%12\relax}%
\providecommand \@@startlink[1]{}%
\providecommand \@@endlink[0]{}%
\providecommand \url  [0]{\begingroup\@sanitize@url \@url }%
\providecommand \@url [1]{\endgroup\@href {#1}{\urlprefix }}%
\providecommand \urlprefix  [0]{URL }%
\providecommand \Eprint [0]{\href }%
\providecommand \doibase [0]{https://doi.org/}%
\providecommand \selectlanguage [0]{\@gobble}%
\providecommand \bibinfo  [0]{\@secondoftwo}%
\providecommand \bibfield  [0]{\@secondoftwo}%
\providecommand \translation [1]{[#1]}%
\providecommand \BibitemOpen [0]{}%
\providecommand \bibitemStop [0]{}%
\providecommand \bibitemNoStop [0]{.\EOS\space}%
\providecommand \EOS [0]{\spacefactor3000\relax}%
\providecommand \BibitemShut  [1]{\csname bibitem#1\endcsname}%
\let\auto@bib@innerbib\@empty
\bibitem [{\citenamefont {{\v{C}}ern{\`y}}(1964)}]{vcerny1964poznamka}%
  \BibitemOpen
  \bibfield  {author} {\bibinfo {author} {\bibfnamefont {J.}~\bibnamefont {{\v{C}}ern{\`y}}},\ }\href@noop {} {\bibfield  {journal} {\bibinfo  {journal} {Matematicko-fyzik{\'a}lny {\v{c}}asopis}\ }\textbf {\bibinfo {volume} {14}},\ \bibinfo {pages} {208} (\bibinfo {year} {1964})}\BibitemShut {NoStop}%
\bibitem [{\citenamefont {J{\"u}rgensen}(2008)}]{jurgensen2008synchronization}%
  \BibitemOpen
  \bibfield  {author} {\bibinfo {author} {\bibfnamefont {H.}~\bibnamefont {J{\"u}rgensen}},\ }\href@noop {} {\bibfield  {journal} {\bibinfo  {journal} {Information and Computation}\ }\textbf {\bibinfo {volume} {206}},\ \bibinfo {pages} {1033} (\bibinfo {year} {2008})}\BibitemShut {NoStop}%
\bibitem [{\citenamefont {Ryzhikov}(2020)}]{ryzhikov2020synchronizing}%
  \BibitemOpen
  \bibfield  {author} {\bibinfo {author} {\bibfnamefont {A.}~\bibnamefont {Ryzhikov}},\ }\emph {\bibinfo {title} {Synchronizing automata and coding theory}},\ \href@noop {} {Ph.D. thesis},\ \bibinfo  {school} {Universit{\'e} Paris-Est} (\bibinfo {year} {2020})\BibitemShut {NoStop}%
\bibitem [{\citenamefont {van Hoorn~BSc}\ \emph {et~al.}(2022)\citenamefont {van Hoorn~BSc}, \citenamefont {Don},\ and\ \citenamefont {Bosma}}]{van2022synchronizing}%
  \BibitemOpen
  \bibfield  {author} {\bibinfo {author} {\bibfnamefont {A.}~\bibnamefont {van Hoorn~BSc}}, \bibinfo {author} {\bibfnamefont {H.}~\bibnamefont {Don}},\ and\ \bibinfo {author} {\bibfnamefont {W.}~\bibnamefont {Bosma}},\ }\href@noop {} {\bibfield  {journal} {\bibinfo  {journal} {System}\ }\textbf {\bibinfo {volume} {1}},\ \bibinfo {pages} {42} (\bibinfo {year} {2022})}\BibitemShut {NoStop}%
\bibitem [{\citenamefont {Eppstein}(1990)}]{eppstein1990reset}%
  \BibitemOpen
  \bibfield  {author} {\bibinfo {author} {\bibfnamefont {D.}~\bibnamefont {Eppstein}},\ }\href@noop {} {\bibfield  {journal} {\bibinfo  {journal} {SIAM Journal on Computing}\ }\textbf {\bibinfo {volume} {19}},\ \bibinfo {pages} {500} (\bibinfo {year} {1990})}\BibitemShut {NoStop}%
\bibitem [{\citenamefont {Grudka}\ \emph {et~al.}(2025)\citenamefont {Grudka}, \citenamefont {Karczewski}, \citenamefont {Kurzy{\'n}ski}, \citenamefont {Stempin}, \citenamefont {W{\'o}jcik},\ and\ \citenamefont {W{\'o}jcik}}]{grudka2025quantum}%
  \BibitemOpen
  \bibfield  {author} {\bibinfo {author} {\bibfnamefont {A.}~\bibnamefont {Grudka}}, \bibinfo {author} {\bibfnamefont {M.}~\bibnamefont {Karczewski}}, \bibinfo {author} {\bibfnamefont {P.}~\bibnamefont {Kurzy{\'n}ski}}, \bibinfo {author} {\bibfnamefont {J.}~\bibnamefont {Stempin}}, \bibinfo {author} {\bibfnamefont {J.}~\bibnamefont {W{\'o}jcik}},\ and\ \bibinfo {author} {\bibfnamefont {A.}~\bibnamefont {W{\'o}jcik}},\ }\href@noop {} {\bibfield  {journal} {\bibinfo  {journal} {arXiv preprint arXiv:2502.09522}\ } (\bibinfo {year} {2025})}\BibitemShut {NoStop}%
\bibitem [{\citenamefont {{Naeem Irfan}}\ \emph {et~al.}(2013)\citenamefont {{Naeem Irfan}}, \citenamefont {Oriat},\ and\ \citenamefont {Groz}}]{NAEEMIRFAN201389}%
  \BibitemOpen
  \bibfield  {author} {\bibinfo {author} {\bibfnamefont {M.}~\bibnamefont {{Naeem Irfan}}}, \bibinfo {author} {\bibfnamefont {C.}~\bibnamefont {Oriat}},\ and\ \bibinfo {author} {\bibfnamefont {R.}~\bibnamefont {Groz}}\ }(\bibinfo  {publisher} {Elsevier},\ \bibinfo {year} {2013})\ pp.\ \bibinfo {pages} {89--139}\BibitemShut {NoStop}%
\bibitem [{\citenamefont {Schumacher}\ and\ \citenamefont {Werner}(2004)}]{schumacher2004reversible}%
  \BibitemOpen
  \bibfield  {author} {\bibinfo {author} {\bibfnamefont {B.}~\bibnamefont {Schumacher}}\ and\ \bibinfo {author} {\bibfnamefont {R.~F.}\ \bibnamefont {Werner}},\ }\href@noop {} {\bibfield  {journal} {\bibinfo  {journal} {arXiv preprint quant-ph/0405174}\ } (\bibinfo {year} {2004})}\BibitemShut {NoStop}%
\bibitem [{\citenamefont {P{\'e}rez-Delgado}\ and\ \citenamefont {Cheung}(2007)}]{perez2007local}%
  \BibitemOpen
  \bibfield  {author} {\bibinfo {author} {\bibfnamefont {C.~A.}\ \bibnamefont {P{\'e}rez-Delgado}}\ and\ \bibinfo {author} {\bibfnamefont {D.}~\bibnamefont {Cheung}},\ }\href@noop {} {\bibfield  {journal} {\bibinfo  {journal} {Physical Review A—Atomic, Molecular, and Optical Physics}\ }\textbf {\bibinfo {volume} {76}},\ \bibinfo {pages} {032320} (\bibinfo {year} {2007})}\BibitemShut {NoStop}%
\bibitem [{\citenamefont {Wiesner}(2008)}]{wiesner2008quantum}%
  \BibitemOpen
  \bibfield  {author} {\bibinfo {author} {\bibfnamefont {K.}~\bibnamefont {Wiesner}},\ }\href@noop {} {\bibfield  {journal} {\bibinfo  {journal} {arXiv preprint arXiv:0808.0679}\ } (\bibinfo {year} {2008})}\BibitemShut {NoStop}%
\bibitem [{\citenamefont {Gr{\"o}ssing}\ and\ \citenamefont {Zeilinger}(1988)}]{grossing1988quantum}%
  \BibitemOpen
  \bibfield  {author} {\bibinfo {author} {\bibfnamefont {G.}~\bibnamefont {Gr{\"o}ssing}}\ and\ \bibinfo {author} {\bibfnamefont {A.}~\bibnamefont {Zeilinger}},\ }\href@noop {} {\bibfield  {journal} {\bibinfo  {journal} {Complex systems}\ }\textbf {\bibinfo {volume} {2}},\ \bibinfo {pages} {197} (\bibinfo {year} {1988})}\BibitemShut {NoStop}%
\bibitem [{\citenamefont {Farrelly}(2020)}]{farrelly2020review}%
  \BibitemOpen
  \bibfield  {author} {\bibinfo {author} {\bibfnamefont {T.}~\bibnamefont {Farrelly}},\ }\href@noop {} {\bibfield  {journal} {\bibinfo  {journal} {Quantum}\ }\textbf {\bibinfo {volume} {4}},\ \bibinfo {pages} {368} (\bibinfo {year} {2020})}\BibitemShut {NoStop}%
\bibitem [{\citenamefont {Moore}\ and\ \citenamefont {Crutchfield}(2000)}]{moore2000quantum}%
  \BibitemOpen
  \bibfield  {author} {\bibinfo {author} {\bibfnamefont {C.}~\bibnamefont {Moore}}\ and\ \bibinfo {author} {\bibfnamefont {J.~P.}\ \bibnamefont {Crutchfield}},\ }\href@noop {} {\bibfield  {journal} {\bibinfo  {journal} {Theoretical Computer Science}\ }\textbf {\bibinfo {volume} {237}},\ \bibinfo {pages} {275} (\bibinfo {year} {2000})}\BibitemShut {NoStop}%
\bibitem [{\citenamefont {Lovett}\ \emph {et~al.}(2010)\citenamefont {Lovett}, \citenamefont {Cooper}, \citenamefont {Everitt}, \citenamefont {Trevers},\ and\ \citenamefont {Kendon}}]{lovett2010universal}%
  \BibitemOpen
  \bibfield  {author} {\bibinfo {author} {\bibfnamefont {N.~B.}\ \bibnamefont {Lovett}}, \bibinfo {author} {\bibfnamefont {S.}~\bibnamefont {Cooper}}, \bibinfo {author} {\bibfnamefont {M.}~\bibnamefont {Everitt}}, \bibinfo {author} {\bibfnamefont {M.}~\bibnamefont {Trevers}},\ and\ \bibinfo {author} {\bibfnamefont {V.}~\bibnamefont {Kendon}},\ }\href@noop {} {\bibfield  {journal} {\bibinfo  {journal} {Physical Review A—Atomic, Molecular, and Optical Physics}\ }\textbf {\bibinfo {volume} {81}},\ \bibinfo {pages} {042330} (\bibinfo {year} {2010})}\BibitemShut {NoStop}%
\bibitem [{\citenamefont {Costa}\ \emph {et~al.}(2018)\citenamefont {Costa}, \citenamefont {Portugal},\ and\ \citenamefont {de~Melo}}]{costa2018quantum}%
  \BibitemOpen
  \bibfield  {author} {\bibinfo {author} {\bibfnamefont {P.~C.}\ \bibnamefont {Costa}}, \bibinfo {author} {\bibfnamefont {R.}~\bibnamefont {Portugal}},\ and\ \bibinfo {author} {\bibfnamefont {F.}~\bibnamefont {de~Melo}},\ }\href@noop {} {\bibfield  {journal} {\bibinfo  {journal} {Quantum Information Processing}\ }\textbf {\bibinfo {volume} {17}},\ \bibinfo {pages} {1} (\bibinfo {year} {2018})}\BibitemShut {NoStop}%
\bibitem [{\citenamefont {Huerta~Alderete}\ \emph {et~al.}(2020)\citenamefont {Huerta~Alderete}, \citenamefont {Singh}, \citenamefont {Nguyen}, \citenamefont {Zhu}, \citenamefont {Balu}, \citenamefont {Monroe}, \citenamefont {Chandrashekar},\ and\ \citenamefont {Linke}}]{huerta2020quantum}%
  \BibitemOpen
  \bibfield  {author} {\bibinfo {author} {\bibfnamefont {C.}~\bibnamefont {Huerta~Alderete}}, \bibinfo {author} {\bibfnamefont {S.}~\bibnamefont {Singh}}, \bibinfo {author} {\bibfnamefont {N.~H.}\ \bibnamefont {Nguyen}}, \bibinfo {author} {\bibfnamefont {D.}~\bibnamefont {Zhu}}, \bibinfo {author} {\bibfnamefont {R.}~\bibnamefont {Balu}}, \bibinfo {author} {\bibfnamefont {C.}~\bibnamefont {Monroe}}, \bibinfo {author} {\bibfnamefont {C.}~\bibnamefont {Chandrashekar}},\ and\ \bibinfo {author} {\bibfnamefont {N.~M.}\ \bibnamefont {Linke}},\ }\href@noop {} {\bibfield  {journal} {\bibinfo  {journal} {Nature communications}\ }\textbf {\bibinfo {volume} {11}},\ \bibinfo {pages} {3720} (\bibinfo {year} {2020})}\BibitemShut {NoStop}%
\bibitem [{\citenamefont {Kempe}(2009)}]{kempe2009quantum}%
  \BibitemOpen
  \bibfield  {author} {\bibinfo {author} {\bibfnamefont {J.}~\bibnamefont {Kempe}},\ }\href@noop {} {\bibfield  {journal} {\bibinfo  {journal} {Contemporary Physics}\ }\textbf {\bibinfo {volume} {50}},\ \bibinfo {pages} {339} (\bibinfo {year} {2009})}\BibitemShut {NoStop}%
\bibitem [{\citenamefont {Nielsen}\ and\ \citenamefont {Chuang}(2010)}]{nielsen2010quantum}%
  \BibitemOpen
  \bibfield  {author} {\bibinfo {author} {\bibfnamefont {M.~A.}\ \bibnamefont {Nielsen}}\ and\ \bibinfo {author} {\bibfnamefont {I.~L.}\ \bibnamefont {Chuang}},\ }\href@noop {} {\emph {\bibinfo {title} {Quantum computation and quantum information}}}\ (\bibinfo  {publisher} {Cambridge university press},\ \bibinfo {year} {2010})\BibitemShut {NoStop}%
\end{thebibliography}%
\end{document}